\documentclass[%
 aip,
 amsmath,amssymb,
 reprint,%
]{revtex4-2}

\usepackage{graphicx}
\usepackage{dcolumn}
\usepackage{bm}

\usepackage[utf8]{inputenc}
\usepackage[T1]{fontenc}
\usepackage{mathptmx}
\usepackage{etoolbox}

\usepackage[hidelinks]{hyperref}
\usepackage{orcidlink}

\makeatletter
\def\@email#1#2{%
 \endgroup
 \patchcmd{\titleblock@produce}
  {\frontmatter@RRAPformat}
  {\frontmatter@RRAPformat{\produce@RRAP{*#1\href{mailto:#2}{#2}}}\frontmatter@RRAPformat}
  {}{}
}%
\makeatother

\begin{document}  
\title{Overcoming experimental obstacles in  two-dimensional spectroscopy of a single molecule}

\author{Sanchayeeta Jana\orcidlink{0009-0002-0383-7351}}

\altaffiliation[Current address: ]{Max-Born-Institut, Max-Born-Str. 2A, 12489 Berlin, Germany}
\affiliation{Experimental Physics III, University of Bayreuth, 95447 Bayreuth, Germany}

\author{Simon Durst}
\affiliation{Experimental Physics III, University of Bayreuth, 95447 Bayreuth, Germany}

\author{Lucas Ludwig\orcidlink{0009-0006-1472-0537}}

\altaffiliation[Current address: ]{DESKO GmbH, Gottlieb-Keim-Str. 56, 95448 Bayreuth, Germany}
\affiliation{Experimental Physics III, University of Bayreuth, 95447 Bayreuth, Germany}

\author{Markus Lippitz\orcidlink{0000-0003-1218-6511}}
\email{markus.lippitz@uni-bayreuth.de}
\affiliation{Experimental Physics III, University of Bayreuth, 95447 Bayreuth, Germany}

\date{\today}

\begin{abstract}
    Two-dimensional electronic spectroscopy provides information on coupling and energy transfer between excited states on ultrafast timescales. Only recently, incoherent fluorescence detection has made it possible to combine this method with single-molecule optical spectroscopy to reach the ultimate limit of sensitivity.  The main obstacle has been the low number of photons detected due to limited photostability. Here we discuss the key experimental choices that allowed us to overcome these obstacles: broadband acousto-optic modulation, accurate phase-locked loops, photon-counting lock-in detection, delay stage linearization, and detector dead-time compensation. We demonstrate how the acquired photon stream data can be used to post-select detection events according to specific criteria.

\end{abstract}

\maketitle

\section{Introduction}

Two-dimensional electronic spectroscopy (2DES) is a well-established ultrafast technique for studying energy transfer dynamics \cite{Jonas2003,Brixner2005,Tiwari2013}. However, most of these experiments are performed on ensembles of molecules. While ensemble studies reveal the collective behavior, single molecule spectroscopy provides insight into the heterogeneity of the system by transforming the ensemble average into a distribution of spectroscopic properties\cite{Adhikari2022,EvanDijk05,Moya2022}.

The most common form of 2D spectroscopy uses three laser pulses to excite the sample. The emitted coherent radiation is then detected after interference with a fourth pulse acting as a local oscillator. Different nonlinear signals are separated by spatial phase matching. This technique requires a sample volume larger than $\lambda^{3}$, so it can't be used for a single-molecule experiment \cite{Tiwari2018}. Action-based 2D spectroscopy, on the other hand, uses a fourth pulse to bring the system into a population state from which an incoherent action signal  is detected, e.g., a fluorescence photon \cite{Tekavec2007}, a photocurrent \cite{bolzonello2021}, or a photoelectron \cite{uhl2021}. Although these two techniques are conceptually similar, they are not identical since the interaction of the fourth pulse with the sample results in additional excitation pathways \cite{kuhn2020, Maly2020, Bolzonello2023, Javed2024}.

Fluorescence-detected two-dimensional electronic spectroscopy (F-2DES) uses four collinear pulses to excite the sample and detects the incoherent fluorescence signal \cite{li_multiphoton_2007, Tekavec2007,mueller_rapid_2019}. Different phase-cycling schemes can be used to extract the phase information from the fluorescence signal \cite{Fuller2015, draeger_rapid-scan_2017, Agathangelou2021, jayachandran_cogwheel_2024}. Because an incoherent fluorescence signal is detected, spatial phase matching and sample volume do not play a role. Thus, this approach can be combined with single-molecule fluorescence spectroscopy to measure the 2D spectra of individual molecules. However, a significant challenge for such an experiment is the limited photostability of the dye molecules at room temperature. In general, most dye molecules emit about $10^{6}$ photons before permanently photobleaching \cite{Herkert_roadmap_21}. Performing a nonlinear experiment with this limited number of photons is challenging.

Recently, we demonstrated 2D spectroscopy of single dibenzoterrylene (DBT) molecules in a PMMA matrix using fluorescence as a reporter \cite{Jana2024}. We have shown that about $10^{5}$ photons are sufficient to measure nonlinear 2D spectra of such a molecule. Here, we discuss the key experimental decisions that led to this result.

The detection scheme must be photon efficient. Rapid phase cycling combined with lock-in detection allows simultaneous measurement of all linear and nonlinear processes. Each photon is used to determine linear absorption, rephasing and non-rephasing spectra, and all other mixing products. This requires acusto-optic modulation of broadband laser pulses and lock-in demodulation of a stream of discrete photodetection events. We will discuss how to compensate for the spectral dispersion of the modulators, how to generate the reference frequency, and how to remove the nonlinearity due to detector dead time.
The measurement sequence must be fast because a fluorescent molecule bleaches within a few tens of seconds. To this end, we have chosen fast mechanical delay stages to generate the pulse sequence. We will describe how the stage motion can be linearized to interferometric accuracy and how other phase effects can be corrected. Finally, the acquired photon stream contains much more information than just the 2D spectrum. We show how the stream can be sliced by the time the molecules spend in the excited state before photoemission.

\section{Experimental Setup}

\subsection{Overview}

\begin{figure*}
    \centering\includegraphics*[]{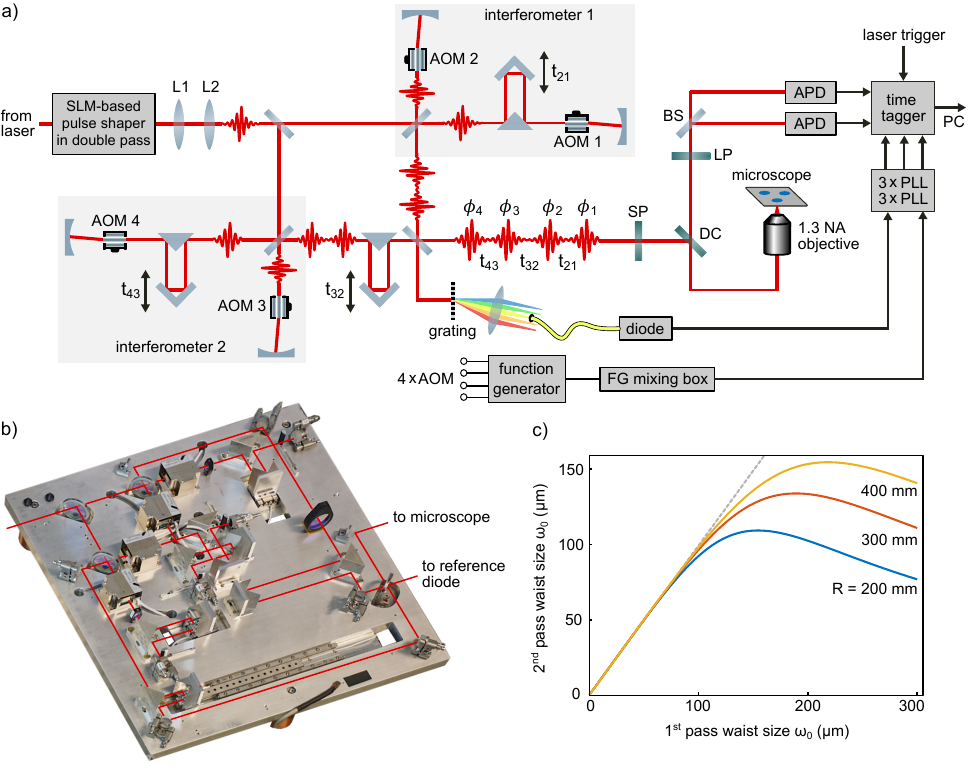}
    \caption{\textbf{Experimental setup} (a) Schematic of the setup: Four collinear pulses are generated using a four-arm cascaded interferometer. Four AOMs phase modulate the pulses by $\phi_{i}$, and three mechanical delay stages control the delay $t_{ij}$ between the pulses. This pulse sequence is directed to a homemade inverted microscope. The red-shifted fluorescence signal from the molecule is registered by two single-photon counting detectors and recorded by a time tagger. The second output of the interferometer is spectrally filtered and used as a reference signal for lock-in detection, where three phase-locked loops (PLLs) are used to determine the relative phase differences $\phi_{ij}$ between the pulse pairs. Another three PLLs are used to determine the phase of the reference signal with respect to the electronically mixed function generator (FG) signal. (b) Image of the interferometer with the beam path superimposed. (c) Relation between the Gaussian focus waist size $\omega_0$ in the first and second pass through the AOM ($\lambda = 750$~nm). A larger radius of curvature of the spherical mirror would allow a larger waist size but also increase the size of the baseplate. }
    \label{fig:setup}
    \end{figure*}

    Our setup follows the design proposed by Tekavec et al. \cite{Tekavec2007} with some modifications. The laser system is a Ti:Sa oscillator (Laser Quantum, Venteon Power) with a repetition rate of 80~MHz and about 20~fs pulse length. After pre-compensation of the group delay dispersion in a pulse shaper in double pass configuration \cite{brinks2011}, we use two Michelson interferometers cascaded in a Mach-Zehnder interferometer to generate the sequence of four pulses (Fig.~\ref{fig:setup}a). Three mechanical delay stages  (Smaract, $2 \times$ SLC-2430, $1\times$ SLC-24150) control the delays between the pulses $t_{21}$, $t_{32}$ and $t_{43}$.

    Rapid phase cycling means that each pulse is phase-modulated with a phase $\phi_{i}(t) = \Omega_i \, t$, $i = 1,\dots,4$. This phase 'tags' each laser field and is propagated through all interactions. The detected fluorescence signal carries this phase information as an amplitude modulation at a frequency that is a characteristic linear combination of the $\Omega_i$. By choosing the detection frequency, we can select the excitation path we observe. For this phase modulation we use in our setup an acousto-optical modulator (AOM) in each arm of the interferometers (G \& H, 30~mm crystalline quartz).
     The deflection angle of an AOM is given by  
    \begin{equation} \label{eq: deflection angle}
        \theta_{D} = \frac{\lambda f}{v} \quad ,
    \end{equation}
    where $f$ is the RF frequency applied to the AOM, $v$ is the acoustic velocity through the AOM crystal, and $\lambda$ is the wavelength of the input beam. Since the deflection angle is wavelength-dependent, different parts of the laser spectrum will be diffracted differently by the AOM. As a result, for a broadband beam, there would be a spatial separation of the spectra at the output of an AOM after a single pass. After 2~m of travel, there would be about 3~mm of separation between the blue and red parts of our spectrum.

However, placing the AOM at the center of curvature of a spherical folding mirror avoids this effect. In a ray optics image, all spectral components are reflected back into themselves. 
In a wave optics image, the wavefront of the Gaussian beam must match the spherical shape of the mirror for perfect reflection, otherwise, the focus waist size $\omega_0$ will be different in the second pass (Fig.~\ref{fig:setup}c). In addition, the AOM efficiency depends on the waist size. In our case, the optimal waist size for the AOM is about 150~\textmu m, resulting in a single-pass efficiency of 78 \%. To keep the overall size of the interferometer small, we have chosen $R=200$~mm as the radius of curvature of the spherical mirror. This results in a slightly smaller waist size ($\approx 100$~\textmu m), which is optimized by the two lenses (L1 \& L2) in front of the interferometer. Our double-pass efficiency is 39~\%, a little lower than the square of the single-pass efficiency. The opening angle of the beam is about 5~mrad, still small enough compared to the deflection angle of 11~mrad. We have chosen quartz as the material for the AOM crystal because it results in a comparatively low group delay dispersion (see table \ref{table:AOM_materials}).

    \begin{table}
        \begin{tabular}{ccccc}
        material & $L$ (mm) & $P$ (W) & $\theta_{D}$ (mrad) & GDD (fs$^2$) \\
        quartz & 30            & 6          &     10.9            & 1~100 \\
        SF57 & 11            & 3          &       18.3          & 2~200 \\
        TeO$_2$ & 20            & 1          &        14.8         & 9~900 \\
        \end{tabular}
        \caption{Comparison of different materials for AOMs. We chose quartz: even though the required length $L$ and power $P$ is largest, the overall group delay dispersion (GDD) is lowest. The deflection angle $\theta_{D}$ is small but sufficient. 
        \label{table:AOM_materials} }
    \end{table}

We designed the interferometer in a 3D modelling software (Autodesk Inventor) and built it over a custom-designed plate (Fig.~\ref{fig:setup}b), inspired by Ref.~\onlinecite{Bristow2009}. The interferometer is relatively compact (baseplate dimensions 35 cm $\times$ 40 cm, 1.5~cm thick). All four arms of the interferometer are perfectly identical. Thermal expansion will thus, to a first approximation, only change the less sensitive 'population time' $t_{32}$ of the Mach-Zehnder interferometer and not the two Michelson interferometers.  To drive the quartz AOMs, we need high RF power ($\simeq 6$~W), which heats up the AOMs and, thus, the baseplate. We use cast aluminum for the baseplate and copper pedestals to mount the baseplate on the optical table for better heat dissipation. We place the interferometer in an acoustic foam box to reduce the effect of ambient vibration on the interferometer and to dampen thermal fluctuations of the environment. The baseplate temperature is approximately 10~K above room temperature.

The interferometer has two outputs. One output is directed to a home-built fluorescence microscope. The light reflected via a dichroic beam splitter (DC) is used to excite the sample using a high NA (1.3) objective (Olympus UPlanFL 100x). The fluorescence signal from the sample is collected by the same objective and detected by two single-photon counting avalanche photodiodes (APD, Excelitas SPCM-AQRH). A short pass (SP) and a long pass (LP) filter are used to suppress laser back reflection. The APD photon pulses are registered using a time tagger (Swabian Instruments Time Tagger 20). 

The second output of the interferometer is used to generate a reference signal for lock-in detection. A grating in the focus of a single-mode fiber coupler selects a narrow spectral range that is detected by a photodiode. This stretches the pulse in the time domain and allows interference over a large delay range. Since the spectral resolution is determined by the length of the delay range, the spectral width of the reference signal essentially determines the ultimate spectral resolution of our instrument. Three phase-locked loops (PLL) in a field-programmable gate array (FPGA)  generate trigger pulses in phase with the beat signals over the three delay stages. This allows phase-sensitive lock-in detection of the fluorescence signal.
    
A second type of reference signal is generated by electronically mixing and low-pass filtering the AOM drive signals. This produces beat signals similar to the interference across the delay stages. Another three PLLs are used to detect these electronically mixed phases and compare them with the optical reference phases. This provides information about the drift and vibration of the interferometer and allows us to interferometrically determine the position of the delay stages.

\subsection{Phase-Locked Loop (PLL)}

The interferometer is only passively stabilized. Fluctuations and drifts in the path length difference are not compensated, but are followed synchronously by the detection electronics. In contrast to the design by  Tekavec et al. \cite{Tekavec2007}, we use only one reference diode that contains the beat signals over all three delays. Three phase-locked loops (PLL) are used to extract the phase information of the reference signal.

\begin{figure*}
        \includegraphics*[]{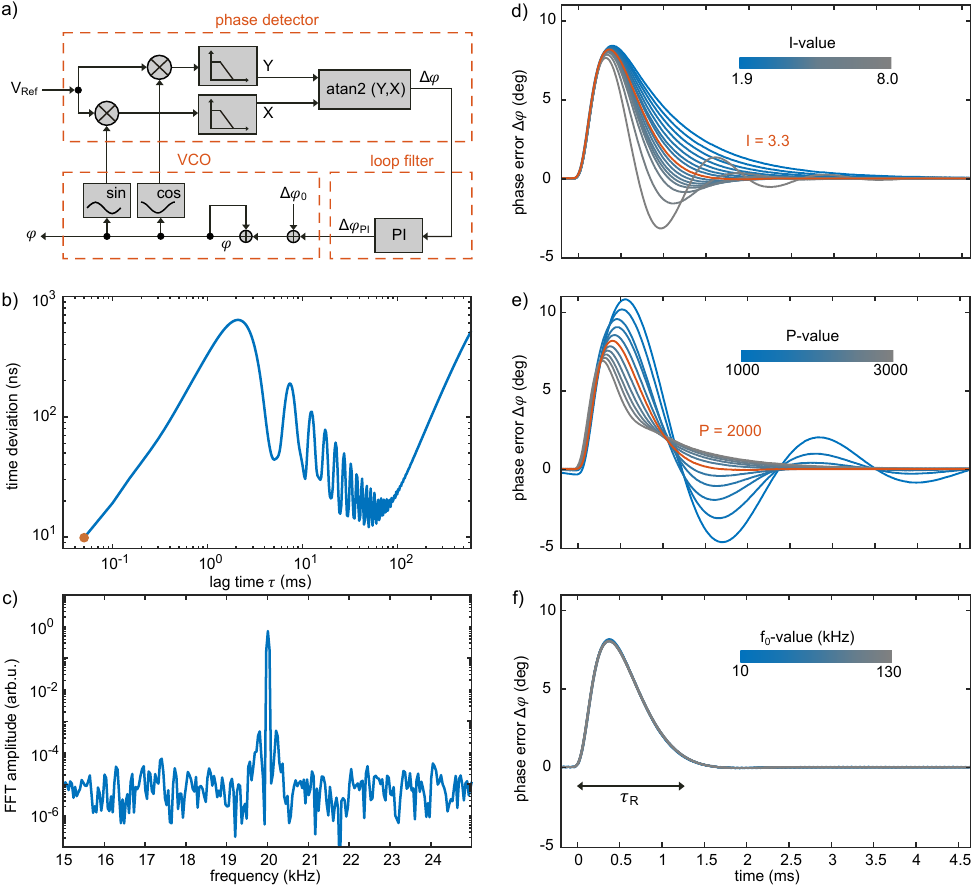}
        \caption{\textbf{Phase-locked loop (PLL)} (a)  Schematic of the PLL. (b) Time deviation $\sigma(\tau)$ of the interferomter for $f_{0}$ = 20 kHz. The oscillation corresponds to a modulation at 200~Hz. (c) The reference signal in the frequency domain. In addition to the 20 kHz signal, sidebands indicate the  200 Hz modulation, most likely due to mechanical oscillations of the interferometer. (d) Step response of the PLL for different $I$ parameters, (e) for different $P$ parameters, and (f) for different $f_{0}$ values.}
        \label{fig_PLL}
\end{figure*}

We implemented a Costas loop \cite{Best_costas_loops} in an FPGA using the Labview programming environment (National Instruments) on an USB-7856R data acquisition board \footnote{Source code available in a publicly accessible repository at \url{https://github.com/Lippitz-Lab/PLL-on-FPGA}}. Figure \ref{fig_PLL}a shows the schematic of the loop, which is repeated at a sampling rate of 1~MHz. First, the phase difference between the input signal and a local oscillator is determined by multiplication and low-pass filtering. This step determines which of the three beat frequencies will be locked. The filter has a cutoff frequency of 4~kHz and a roll-off of 24~dB/oct. A notch filter (4~kHz width) helps to suppress the closest of the other beat signals. The phase difference is computed by a numerical atan2 function that takes into account the signs of the arguments and thus maps to a full $2\pi$ interval. The second component is the loop filter, which we implement as a proportional-integral (PI) feedback controller. It determines the response of the PLL to varying input signals. The third component is the numerical analog of a voltage controlled oscillator (VCO). We increment the phase of the local oscillator at each iteration of the 1~MHz acquisition loop by a fixed value $\Delta \phi_0$ equal to the nominal frequency of the reference signal, rolling over at $2\pi$. The output of the PI controller additionally modifies this phase increment within limits corresponding to $\pm 1$~kHz frequency. We close the loop by calculating the sine and cosine of the phase. This implementation has the advantage that the loop locks at zero phase difference regardless of the start condition.

With optimized PI parameters (see below) we recover the phase of the beat signals once per microsecond. For the time tagger, we generate a trigger pulse when the phase crosses zero. To reduce the jitter to less than 1~\textmu s, we linearly interpolate successive phase values in a faster digital loop to a 5~ns time base. The time delay between the zero crossing of the phase at the analog input and the generated trigger adds to the overall phase response of all other electronics and is zeroed during alignment (see below).

The response of the PLL depends on the $P$ and $I$ parameters. To find the optimum values, we use a function generator as PLL input. We generate a sinusoidal signal with a frequency of 20~kHz, frequency modulated by shift of 100~Hz in a square wave. We set the nominal frequency of the PLL to $f_{0} = 20$~kHz and measure the phase response of the PLL for different $I$ parameters keeping $P = 2000$ constant and for different $P$ parameters keeping $I = 3.3$ constant. (The values given have arbitrary units) The response time $\tau_{R}$ of the PLL is the time it takes for the PLL to lock to the input signal, or in other words, the time it takes for the phase difference $\Delta \varphi$ to relax to zero. As expected for a PI-controller, we observe that as the $I$ value increases, the response time of the PLL decreases, but at higher $I$ values the phase response becomes oscillatory and the response time increases again (Fig. \ref{fig_PLL}d). A similar type of phase response is observed for different $P$ values (Fig. \ref{fig_PLL}e). From these measurements, we find the optimal value for our PLL as $P = 2000$ and $I = 3.3$, which corresponds to the smallest response time of about 1.35~ms.  This means that our PLL can follow all acoustic vibrations of a few 100~Hz frequency, independent of the nominal PLL frequency in the range between 10~kHz and 130~kHz (Fig. \ref{fig_PLL}f). We chose the three beat frequencies from this range as $\Omega_{21} = 20$~kHz, $\Omega_{32} = 68$~kHz, and $\Omega_{43} = 32$~kHz. The corresponding RF frequencies of the AOMs are $78.000$ MHz, $78.010$ MHz, $78.044$ MHz, and $78.060$ MHz, as the double-pass configuration doubles the frequency modulation.

How accurately does the PLL work? How well do the trigger pulses reflect the zero crossing of the beat signal? These questions are answered by the time deviation, which is a variant of the modified Allan deviation \cite{nist_handbook}. Using the software of the time-tagger we compute from all triples of consecutive trigger events $t_{i-1}$, $t_{i}$, $t_{i + 1}$ that are separated by one trigger period $\tau_0$ the time deviation $\sigma$ as
\begin{equation}
    \sigma(\tau_0)^2 = \frac{1}{6}  \, \left<  \left(  t_{i-1} - 2 t_i + t_{i+1} \right)^2 \right> \quad ,
\end{equation}
where the brackets symbolize the average over a large set of trigger events. Longer delays $\tau = n \tau_0$ include an average over $n$ overlapping intervals  \cite{nist_handbook}. Fig. \ref{fig_PLL}b shows an example with a nominal PLL frequency of $\Omega_{21} = 20$ kHz. The shortest time delay $\tau$ is thus 50~\textmu s. We find a time deviation $\sigma(50 \text{ \textmu s}) = 10$~ns, far below the sampling rate of 1~MHz of the PLL. This time deviation corresponds to a phase deviation of less than $0.1^\circ$ at 20~kHz. 
For longer delay times, we find an increasing time deviation with the characteristic pattern of a 200~Hz oscillation, which is also observed in the Fourier transform of the unprocessed reference diode signal (Fig.~\ref{fig_PLL}c). It is most likely a mechanical oscillation of the interferometer, which would lead to a phase deviation of up to $10^\circ$, but the PLL follows this oscillation. At even longer times, we observe an increase due to the slow drifts of the interferometer, which are also followed by the PLL.

\section{Data processing}

\subsection{Software Lock-In-Detection}

Demodulation of the fluorescence signal at $\Omega_{21}$, $\Omega_{32}$, and $\Omega_{43}$ yields the linear signal contribution due to the interaction of fields 1 \& 2,  2 \& 3, and  3 \& 4, respectively. To recover any nonlinear signal, demodulation must be performed at the appropriate mixing frequencies. For example, demodulation at $\Omega_{21}+\Omega_{43}$ yields the non-rephasing contribution, while demodulation at $\Omega_{21}-\Omega_{43}$ yields the rephasing contribution. A 2D Fourier transform of the demodulated interferogram along the delay $t_{21}$ and $t_{43}$ gives the 2D spectra in the spectral domain.

In a commercial lock-in detector, the input signal is electronically mixed with the reference signal to obtain the in-phase ($x$) and quadrature ($y$) components. However, this scheme can't be used with a photon counter. Instead, we use the time tagger to record the arrival time of each photon $t_k$ and calculate by linear interpolation the phase $\theta_{k} = \theta(t_k)$ of the reference signal at the time of the photon detection.
From this we calculate the complex lock-in signal as 
\begin{equation}
    z = x + i y = \frac{2}{N} \sum_{k=1}^{N} \cos \theta_{k} - i \sin \theta_{k} \quad ,
\end{equation}
where the summation is taken over   $N$ photons. 
We limit the integration domain so that we integrate over full $2\pi$ periods of the reference signal, i.e. 
\begin{equation}
    \theta_{N} \le \theta_1 + m \, 2 \pi < \theta_{N+1}
\end{equation}
with the largest possible integer $m$. Considering all photons, this is a small correction since the modulation period is about a factor of 100 faster than the pixel dwell time. Our approach is based on individual detection events rather than averaged count rates as used by Uhl et al.\cite{uhl2021}, which avoids handling large sparse datasets. More details on the lock-in detection are given in the supplementary material of Ref.~\onlinecite{Jana2024}.

\subsection{Phase correction}

\begin{figure}
        \includegraphics[]{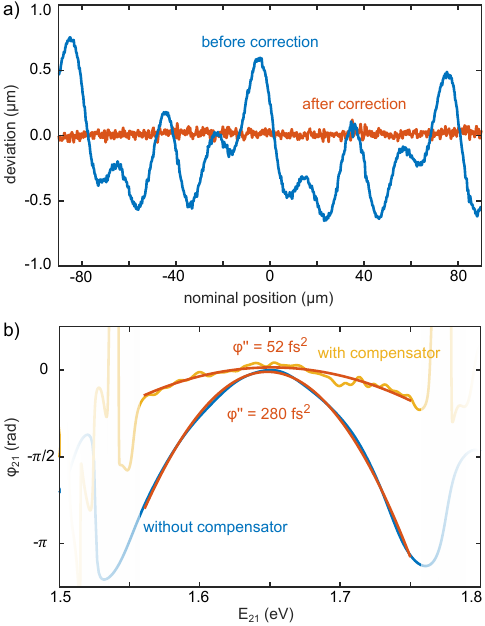}
        \caption{\textbf{Phase correction} (a) Due to deviations of the sensor of the delay stage, the stage goes to a 'wrong' position. However, we can easily correct it after interferometrically  measuring the deviation  once. (b) Spectral phase of the two-arm interferometer with and without compensator plate.}
        \label{fig:Phase_correction}
\end{figure}

We use closed-loop slip-stick stages (Smaract) for the three delays. The resolution of the position sensor is specified as 1~nm and the unidirectional repeatability as $\pm 40$~nm. Important for the experiment is the deviation between real and measured position, i.e. the linearity of the stage. The interferometer allows us to measure the stage position independently. For this purpose, we compare the phase of the optical reference signal with that of the electronic reference (taking into account the factor of two due to the AOM double-pass) and convert the phases into spatial distances using the central wavelength of the reference diode. We find periodic and very reproducible deviations of about $\pm 500$~nm between the expected and realized stage position (Fig.~\ref{fig:Phase_correction}a). One source is the periodic error of the position sensor. Another source is the pitch and yaw rotation of the stage in combination with the approximately 9.5~mm beam height above the stage. We use a look-up table to pre-compensate for these deviations, which drastically reduces the stage non-linearity.

Once the stages are linearized, we measure three 1D interferograms along the three delay stages, calculate the spectral phases $\phi_{ij} (\omega_{ij})$, and fit a third order polynomial \cite{Agathangelou2021}. The constant term gives the phase offset at the reference wavelength. It is set to zero to make the spectrum real and positive. The linear term corresponds to a time shift. We use it to fine tune the zero position of the delay stages. Thus, the constant and linear terms are set to zero by the phase correction terms obtained during alignment.
The quadratic term in the spectral phase corresponds to the group delay dispersion (GDD) difference between the transmitted and reflected beam due to the finite thickness of the beam splitter. To compensate for this GDD mismatch, we place three compensator plates of appropriate thickness in the three arms of the interferometer. As shown in Figure \ref{fig:Phase_correction}b, the GDD difference between the two arms becomes negligible after the addition of the compensator plate.

\subsection{Detector nonlinearity}
\begin{figure*}
    \includegraphics*[]{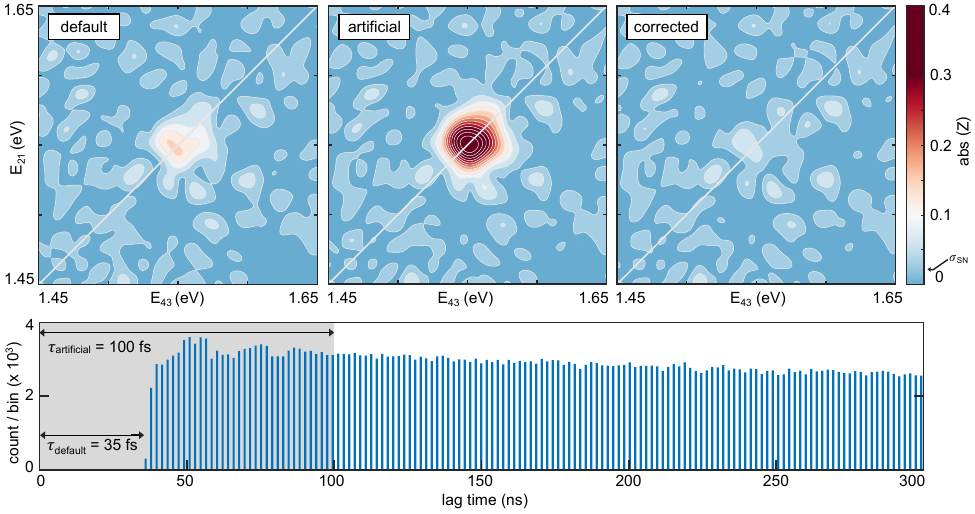}
    \caption{\textbf{Influence of detector nonlinearity on the 2D spectra} The  rephasing 2D spectrum measured with continuous-wave laser reflection at $8 \times 10^{5}$ cps shows some signal (default). This nonlinear signal occurs due to the finite dead time of the detector. Our detector has a dead time of $\tau_{d} =35$~ns (bottom). To demonstrate the dependence of the nonlinear signal on $\tau_{d}$, we artificially increase $\tau_{d}$ to 100 ns and the corresponding rephasing spectra shows an almost 3 times higher signal amplitude (artificial). Using equation (\ref{eq:APD_rephasing}) we could remove the detector nonlinearity (corrected). The arrow in the color bar of the top panel indicates one standard deviation due  to photon shot noise. }
    \label{fig:APD_NL}
\end{figure*}

Care must be taken when dealing with a nonlinear signal because the detector may operate in a nonlinear regime and thus "pollute" the measured nonlinear response of the sample. Single photon detectors have an intrinsic nonlinearity due to their dead time: the time it takes to get ready to detect a second photon after a first photon has been detected. If a photon arrives within this dead time, the detector will not respond, which contributes to a pseudo-nonlinear signal. To test this concept, we measure 2D spectra of continuous-wave laser reflection at $8 \times 10^{5}$~cps (counts per second) on an otherwise blank sample. Ideally, one would detect no nonlinear signal at the demodulation frequencies $\Omega_{21} \pm \Omega_{43}$.
 However, as shown in Fig.\ref{fig:APD_NL} (top panel, "default"), we observe some nonlinear signal in the rephasing spectra and a similar observation was also made for the non-rephasing case. 
 
 Our detector has a dead time of 35~ns, as shown in the histogram in Figure \ref{fig:APD_NL} (bottom panel). To test the influence of the detector dead time on the nonlinear signal, we artificially increase the dead time by removing all photons that come earlier than 100~ns after a preceding photon in our photon stream data. We recompute the rephasing spectra with the artificially increased dead time (Figure \ref{fig:APD_NL} top panel, "artificial"). As expected, the nonlinear signal amplitude increases to almost 3 times that of the "default" scenario.

Since the photon stream follows Poisson statistics on short time scales, we can calculate the detector's contribution to the nonlinear signal from the detector's dead time and the measured linear signal. The probability $P$ of a delay $\tau$ between two successive photons at an intensity $I$ in units of photons per second is
\begin{equation}
    P (\tau) = I e^{-I\tau} \quad \text{with} \quad \int_0^\infty P(\tau) \, d\tau = 1 \quad .
\end{equation}
The probability of a photon within the dead time $\tau_{d}$ is 
\begin{equation}
    \int_{0}^{\tau_{d}} P(\tau) d\tau = 1- I e^{-I \tau_{d}} \quad .
\end{equation}  
So the efficiency $\eta$ of the detector can be defined as 
\begin{equation}
    \eta (I) = 1 - \int_{0}^{\tau_{d}} P(\tau) d\tau = Ie^{-I\tau_{d}} \simeq 1 - I\tau_{d} + \frac{(I\tau_{d})^{2}}{2} + \cdots 
\end{equation}
and the effective measured intensity, $I_\text{eff}$, in counts per second is 
\begin{equation}
    I_\text{eff} = I \cdot \eta (I) = I - I^{2}\tau_{d} + \frac{I^{3}\tau_{d}^{2}}{2} + \cdots \quad ,
\end{equation}
i.e., the detector's contribution to the nonlinear signal up to the first order is $I^{2} \tau_{d}$. This squaring leads to the mixing of the linear signals at $\Omega_{21}$ and $\Omega_{43}$.

In the presence of only linear modulation of amplitude $z_c$ at the frequencies $\Omega_{21}$ and $\Omega_{43}$, the intensity $I$ is given by 
\begin{equation}
    I(t) = I_0 + \frac{1}{2} \sum_{c= (21), (43)} \left( z_c \, e^{i \Omega_c t} + z_c^\star \, e^{-i \Omega_c t}   \right) \quad .
\end{equation}
This gives us at the non-rephasing frequency $\Omega_{21}+ \Omega_{43}$ a detector contribution of 
\begin{equation}
D_{NR} = - 2 \, z_{21} \, z_{43} \, \tau_{d}  \quad .
\end{equation}
Similarly, the detector contribution to the rephasing signal is 
\begin{equation}\label{eq:APD_rephasing}
    D_{R} = - 2 \, z_{21} \, z_{43}^{*} \, \tau_{d}  \quad ,
\end{equation}
where the asterisk indicates the complex conjugate. Thus, we can easily calculate the nonlinear signal from the measured linear signal and subtract the detector's nonlinear contribution.  
When doing so, the rephasing spectra contain only background noise (Figure \ref{fig:APD_NL} top panel "corrected").

Detector nonlinearity becomes relevant when the probability of missing a photon due to dead time approaches the ratio of nonlinear to linear signal amplitude. The latter is a few percent, so photon count rates above about $10^5$ counts per second become critical for our detector (Excelitas SPCM-AQRH). These intensities are high for single molecule fluorescence, but we apply the dead time correction anyway.

\section{Exemplary Results}

\begin{figure*}
    \includegraphics*[]{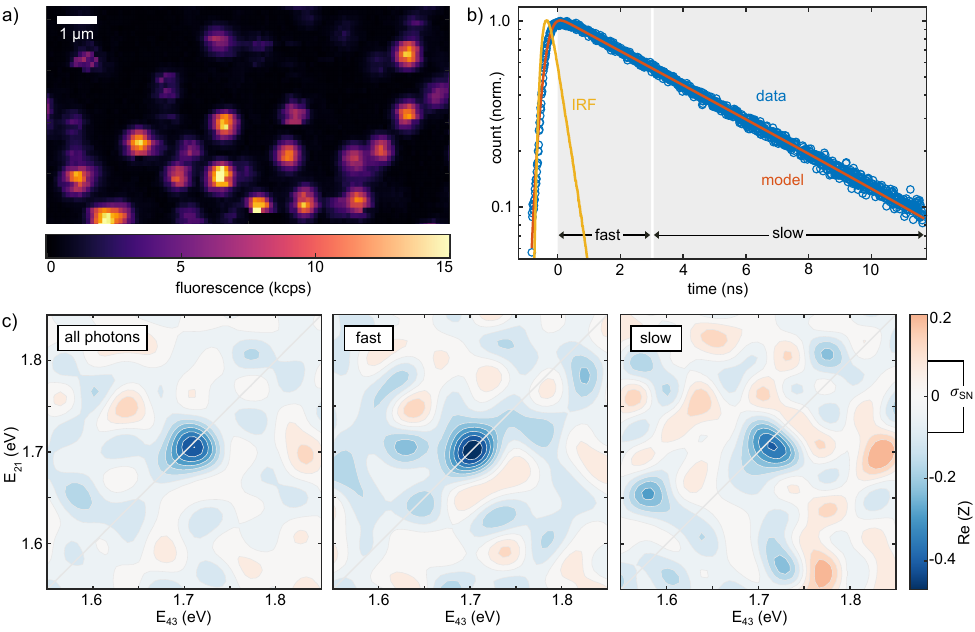}
    \caption{\textbf{2D spectra of single molecules} (a) Exemplary  fluorescence map of DBT molecules in a PMMA thin film. Each bright spot corresponds to a single DBT molecule. (b) TCSPC trace of one such molecule along with the instrument response function (IRF). (c) We sort the photons in two categories: 'fast' ($\tau < 3$~ns) and 'slow' ($\tau >3$~ns). We calculate purely absorptive 2D spectra for each case along with the  'all photons' case. The error bar corresponds to the shot noise level for the latter. The noise level will be $\sqrt{2}$ times higher for the 'fast' and 'slow' case. }
    \label{fig:2DSM}
\end{figure*}

We spin-coat a diluted dibenzoterrylene (DBT) molecule with polymethyl methacrylate (PMMA) in chlorobenzene solution over a clean quartz microscope slide. For more details on sample preparation, see the supplementary material of Ref.~\onlinecite{Jana2024}. By scanning the sample through the laser focus, we obtain a fluorescence map ( Fig.~\ref{fig:2DSM}a ). Each bright spot corresponds to a single DBT molecule. We place the laser focus on one of these molecules and measure 2D spectra by scanning the $t_{21}$ and $t_{43}$ delays in 7 steps of about 10~fs each. We keep a fixed $t_{32}= 50$~fs. To obtain the best signal-to-noise ratio, we repeatedly scan the $t_{21}$ and $t_{43}$ delays until the molecule photobleaches. After summing these repeated measurements, we zero-pad the data and perform a 2D Fourier transform along the two delays. The purely absorptive 2D spectrum $PA$ is defined as the average of nonrephasing ($Z_{NR}$) and rephasing ($Z_{R}$)  spectra after changing the frequency axis of the rephasing spectrum, i.e.,
\begin{equation}
    PA (\omega_{21}, \omega_{43}) = \frac{1}{2} \left[Z_{NR}(\omega_{21}, \omega_{43})+Z_{R}(\omega_{21}, -\omega_{43})\right] \quad .
\end{equation}
 In Fig.~\ref{fig:2DSM}c marked as 'all photons', we show the real part of $PA$ of one DBT molecule. For this molecule, we recorded a total of about $1.27 \times 10^{6}$  photons at an average count rate of $9.86 \times 10^{3}$~cps. The peak amplitude of the nonlinear purely absorptive signal is $-0.47$ which is roughly five times the noise level $\sigma_{SN}$ determined by the photon shot noise.

One advantage of our technique is that we capture the raw photon stream, which contains a wealth of information. One example is the time the molecule spends in the excited state between excitation and emission. A more complex molecule than DBT may have two or more emitting states with different excited state lifetimes \cite{Moya2022}. Thus, photons detected early after excitation would be predominantly from one state, and late photons would be predominantly from the other state. Sorting the photon stream data by arrival time after excitation could distinguish these states and give better insight into the energy transfer mechanisms.
 
To demonstrate this concept, we plot a histogram of the delay $\tau$ between photo detection and the previous laser pulse, the so-called time-correlated single photon counting (TCSPC) trace in Fig.~\ref{fig:2DSM}b. We observe a single exponential decay corresponding to an excited state lifetime of 5~ns. We split the photon stream into two parts: 'fast' and 'slow', corresponding to delays $\tau$ smaller and larger than 3~ns, respectively.  We choose this cut-off to obtain an almost equal number of photons in both parts. For each part, we compute the 2D spectra separately (Fig.~\ref{fig:2DSM}c). These two spectra would reflect the difference in the emitting state.
For our system, we do not observe any difference in the 2D line shape between the 'fast' and 'slow' emission times. Only the noise level changes proportionally to $1/\sqrt{N_\text{photons}}$, since in our definition the nonlinear signal amplitude is independent of the total number of photons detected. The difference in signal amplitudes between the two spectra lies within this increased noise level.

\section{Conclusion}

Two-dimensional electronic spectroscopy of a single molecule is possible \cite{Jana2024}, but challenging. 
In this article we have discussed the key experimental problems that need to be solved. First of all, no fluorescence photon should be wasted. Rapid phase cycling together with lock-in detection allows to measure all mixing products simultaneously. To this end, we presented the double-pass configuration for spectrally broadband acousto-optic modulation and an accurate three-channel phase-locked loop for phase recovery. Second, the three slip-stick stages allow almost random access to the three temporal delays, but care must be taken to achieve the required linearity of the position sensor. Finally, artifacts in spectral phase and detector nonlinearity must be removed. Both can be done by appropriate correction schemes. Since the experiment is based on photon counting, all statistical methods of single molecule data analysis \cite{Lippitz2005} can be applied. We have demonstrated the slicing of the stream by photon arrival time to potentially distinguish different emitting states. We are confident that such data analysis in combination with an optimized sampling scheme \cite{Roeding2017, Bolzonello2024} and excitation of the molecule by surface plasmon \cite{schorner2020} will open the door to a wide range of systems, from photosynthesis to plasmonic strong coupling.

\section*{Acknowledgements}

 We thank M.~Theisen, M.~Heindl, and C.~Schnupfhagn for their work on the pulse shaper,  F.~Paul for his help in the early stages of this experiment,  and R.~Weiner for the electronics. We acknowledge the financial support of the German Science Foundation (DFG) via IRTG OPTEXC and project 524294906.

 \section*{Author declarations}

 \subsection*{Conflict of Interest} The authors have no conflicts to disclose.

 \subsection*{Author Contributions}

\textbf{Sanchayeeta Jana:} Investigation (equal),  Visualization (equal), Writing -- original draft, Writing -- review \& editing (equal).
                    \textbf{Simon Durst:}  Investigation (equal),  Software (equal), Visualization (equal), Writing -- review \& editing (equal).
                    \textbf{Lucas Ludwig:} Software (equal).
                    \textbf{Markus Lippitz:} Conceptualization, Software (equal), Writing -- review \& editing (equal).

\section*{Data Availability} The data that support the findings of this study are available from the corresponding author upon reasonable request.


\section*{References}
\bibliography{references}

\end{document}